\documentclass[12pt]{article}
\usepackage{epsfig}
\begin{document}
\title{Optical Spin Orientation in Strained Superlattices}

\author{{A.V. Subashiev, L.G. Gerchikov, and A.I. Ipatov}\\
{\em Department of Experimental Physics, State
 Polytechnical University,} \\ {\em 195251,
Polytechnicheskaya 29, St.Petersburg, Russia}}

\maketitle

\date{today}

\begin{abstract}
Optical orientation in the strained semiconductor superlattices is investigated
theoretically. The dependence of the features in spin-polarization spectra on the
structure parameters is clarified. The value of polarization in the first polarization
maximum in the SL structures is shown to grow with the splitting between the hh- and
lh- states of the valence band, the joint strain and confinement effects on the hh1-
lh1 splitting being strongly influenced by the tunneling in the barriers. In strained
structures with high barriers for the holes initial polarization can exceed 95 \%.
Calculated polarization spectra are close to the experimental spectra of polarized
electron emission.
\end{abstract}

\begin{center}
\end{center}

\section*{Introduction}

Strained GaAs-based layers and superlattices are known to be most
effective as photoemitters of spin-polarized electron beams which
are required for high-energy experiments at many electron
accelerator facilities \cite{Slac,JLab,Mami}. Emitters of
spin-polarized electrons will also be essential for future linear
electron colliders (e.g. see Ref. \cite{SPIN2000} for the
references).

High electron polarization is obtained by a spin optical
orientation \cite{OptOr} which relies on the spin-orbit
interaction in the valence band and selection rules for the
interband optical transitions. Electron emission in vacuum at the
near-band edge optical excitation is provided via activation of
the structure surface by Cs(O) co-deposition to obtain a negative
electron affinity.

In structures with compressively strained GaAs layer uniaxial
deformation caused by the lattice mismatch of the working layer
and a substrate results in a splitting of the 4-fold degenerate
$\Gamma_8$ valence band state into only 2-fold spin-degenerate
states: the $\Gamma_6$ (heavy-hole) and the $\Gamma_7$
(light-hole) subband states, the heavy-hole band being moved up.
The electrons excited by a circularly polarized light at the
absorption edge from the heavy-hole subband with angular momentum
$J=3/2$ populate only one electronic spin state of the conduction
band. Therefore the electron polarization in the conduction band
$|P|=100$ \%. At higher energies the light holes with $J=1/2$
start to contribute to the absorption via transitions to the
second electron spin state with opposite direction of the
electron spin orientation. This results in a decrease of the
polarization with the excitation energy above the light-hole band
excitation edge.

Unfortunately, since the critical thickness of coherent growth is
of the order of 10 nm, the photocathode strained layer structures
are usually partly relaxed and thus have rather poor structural
quality. Achievable valence band splittings in these structures
are less than 50 meV \cite{Mair} which results in sizable
polarization losses in the near band-gap excitation \cite{SuKal}
via various processes that lead to the light-hole state
contribution to the optical transition at the absorption edge.

In Quantum Well (QW) or Superlattice (SL) structures the valence
band splitting is caused by the hole confinement in the QW
layers. For these structures the theoretical limit for the
initial polarization at the excitation is also 100 \%. However,
experimentally obtained values of the emitted electron
polarization for unstrained SLs are typically less than 80 \%.
Since experimental studies of the electron depolarization in
transport to the surface and emission in vacuum do not give more
than 10 \% depolarization for high quality thin-layer structures
\cite{Kurt}, the initial electron polarization becomes an
important issue.

The factors limiting polarization values in SL structures are the
heavy-hole-light-hole state mixing for finite hole momenta in the
layer plane and also the hole scattering between minibands which
lead to population of the second spin state.\cite{ADA} To reduce
the effects of these factors the enlarged value of the valence
band splitting is again essential.

The main advantage of the strained SL structures is the
potentially lager valence band splitting that results from the
additive effects of hole confinement and strain. Besides, in the
strained SLs higher total critical thicknesses can be achieved
which ensures homogeneous strain and better structural quality of
the working layers. As a result experimental values close to $P$
= 90 \% were recently reported \cite{MamCle,Nakani}. However, the
band offset in the conduction band is usually larger than in the
valence band, so hole confinement may be accompanied by strong
electronic confinement. In this case the electrons are not mobile
enough to reach the surface without some depolarization.

The emitted electron polarization depends on a number of
structure parameters, including the band offsets, the
transparency of the barrier layers for the electrons and holes
etc. The flexibility of the SL structures allows one to adjust
these parameters, though a specific structure design is needed.
For this purpose it is important to have a qualitative
comprehension of the main factors contributing to the variation
of the polarization spectra with the SL parameters.

Former theoretical studies of the electron optical orientation
were restricted to QW structures \cite{CHerm,Merk} and SL
structures with a fixed set of parameters \cite{ADA}, so that
quantitative predictions for maximum $P$ values in SLs with
different structure parameters were not reliable.

In this paper we concentrate on elucidation of the general
aspects of optical orientation in unstrained and strained SLs and
its dependence on the structure parameters, having in mind the
structure optimization for photocathode applications. The role of
the heterointerface effects on the polarization spectra is
examined. The maximum polarization of electrons at the moment of
excitation is discussed.

\section*{Optical absorption and spin orientation}
\subsection*{Effective  Hamiltonian and band structure calculations}
The starting point of our calculations is finding the band structure and
optical matrix elements for the circularly polarized light excitation and
their dependence on the SL parameters. We use the envelope function
approximation in the multi-band Kane model including the conduction band $
\Gamma_6$, the states of light and heavy holes of the valence band $\Gamma_8$
and also the states of the spin-orbit splitted band $\Gamma_7$.
The electronic polarization defined by the concentration of the
electrons with the spin parallel ($n_\uparrow$) and
antiparallel ($n_\downarrow$) to the light propagation direction as
\begin{equation}
P=(n_\uparrow-n_\downarrow) /(n_\uparrow+n_\downarrow),
\label{P}
\end{equation}
depends on the difference in the absorption for the two sets of states
and therefore is more sensitive to the spectrum details than just an
absorption coefficient.

For this reason we use the approach of Baraff and Gershoni\cite{Bar}, which is based
on expansion of the envelope function in a Fourier series. The advantage of this is
the possibility to trace easily the role of various factors in the electron optical
orientation including the choice of the boundary conditions.

The wave function of the carrier in the SL structure is taken
in the form
\begin{equation}
\Psi ({\bf r})=\sum_{\nu =1}^{N}\psi ^{\nu }({\bf r})u_{\nu}({\bf r})
\end{equation}
where $\nu$ labels the periodic Bloch wave functions
$u_{\nu}(\bf{r})$ of the included bands at the center of the Brillouin
zone (which are supposed to be equal in both layers of the SL).
The set of Bloch functions is taken as $|S\uparrow \rangle $ for the
conduction band and $|X\uparrow \rangle $, $|Y\uparrow \rangle $,
$|Z\uparrow \rangle$ for the valence band and also
their time-reversed conjugates. In this basis the
Hamiltonian is  $8\times 8$ matrix:
\begin{equation}
H=\left(
\begin{array}{cc}
G & \Gamma  \\
-\Gamma ^{*} & G^{*}
\end{array}
\right)   \label{H-r}
\end{equation}
where $G$ and $\Gamma $ are both $4\times 4$ matrixes defined in
Ref. \cite{Kane} and given explicitly in the Appendix.
The band edge values $E_{c,v}$,
the spin-orbital valence band splitting $\Delta $, Kane matrix
elements $P$, and the band effective mass parameters
$A,L,M,N$  are taken to change abruptly at the interfaces.
Inside the $i$-th SL layer, ($i=1$ for a quantum well at $0<z<a$
and $i$= 2 for a barrier at $a<z<a+b$) these parameters are equal
to their corresponding values for the bulk materials.

The SL translational symmetry leads to the Bloch type structure
for the envelopes. Therefore, the states in the SL are characterized by the
carrier wave vector parallel to interface $\mathbf{k}$ and by the component
$q$ normal to interface, while the periodic part of the envelope can be
found as a series expansion of plane waves
\begin{equation}
\psi ^{(\nu )}_{{\bf k},q} ({\bf r}) = \frac{1}{\sqrt{d}}e^{i{\bf k \rho }%
+iqz}\sum_{n}\exp (i\frac{2\pi}{d}nz)A_{n,\nu }({\bf k},q),
\label{Basis}
\end{equation}
where $d=a+b$ is the SL period, and $n = 0,\pm 1, \pm 2...$.
The energy spectrum of the SL as well as the eigenstates of the problem are
found from the solution of the matrix equation
\begin{equation}
\sum_{n\nu }\left\langle m\mu \left| \hat{H}\right| n\nu \right\rangle
A_{n\nu }(\mathbf{k}, q) = \varepsilon (\mathbf{k},q)A_{m\mu }(\mathbf{k},
q).  \label{Schred}
\end{equation}
The strain effects are readily incorporated by including an
additional term, $H_{strain}$, in the effective Hamiltonian
of each layer \cite{Bar}.
\begin{figure}
  \centering
  \epsfig{file=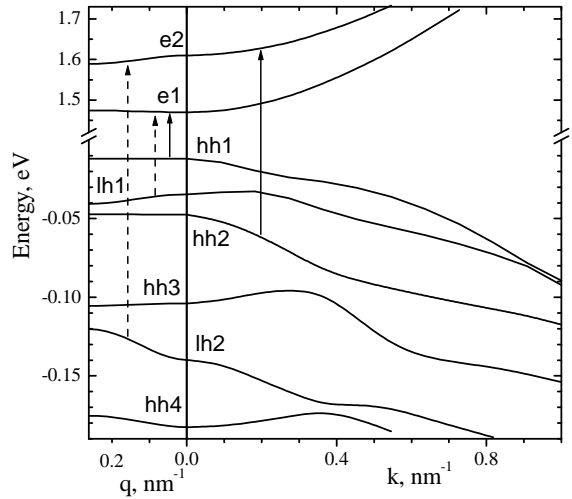,width=7.5cm}
  \caption{Miniband spectrum of GaAs/Al$_{0.4}$Ga$_{0.6}$As 8$\times$
  4 nm superlattice and the major optical transitions to $|\uparrow
\rangle $ (solid) and $|\downarrow \rangle $ states (dashed
arrows) under circularly polarized
light excitation.}\label{FIG1}
\end{figure}
The calculated miniband spectrum is shown in Fig. 1 for a (100)-
oriented lattice-matched  GaAs/Al$_{0.4}$Ga$_{0.6}$As SL
structure assuming the electron barrier height $\Lambda_c $ = 0.313 eV
and the hole barrier $\Lambda_v $ = 0.213 eV. In calculations we
use the material parameters from Ref. \cite{Ram-Moh}
The thickness of the GaAs layer, $a$=8 nm, is chosen to obtain several
miniband states localized in the GaAs well layer  both electrons
and holes, and consequently several optical transitions contributing
to the interband absorption spectrum near the edge. The width of the
barriers are taken to be $b = 4$ nm, so that the barriers are
sufficiently transparent for electrons and light holes.

For the chosen well depth and thickness there are two electron
(e), two light hole (lh) and five heavy hole (hh) confined
quantum levels. The effects of tunnelling through the barriers
are manifested in the electronic and light-hole miniband
dispersion along the SL axis. We note here a small separation
between the second heavy-hole (hh2) and the first light-hole
(lh1) minibands at the center of the Brillouin zone that is
typical for these structures since the effective mass ratio in
GaAs is $m_{hh}/m_{lh}\approx$5 and the lh1 miniband is shifted
up towards the hh2 miniband due to the barrier transparency.

\subsection*{Boundary conditions}
The choice of boundary conditions for the envelope functions
$\psi ^{\nu }$ is a central point for the approaches based on the
effective mass model \cite{Smith}. In the transfer matrix
approach \cite{IvPi,GelMo} a transfer matrix connecting the
components of the envelope function and their derivatives on each
side of an interface is postulated as a characteristic of the
interface. An equivalent approach adopted here specifies variation
of the effective mass Hamiltonian $\hat H$ of the SL given
by Eq. (\ref{H-r}) in the interface region \cite{Bar} by inclusion
additional interface terms \cite{Ivch,Ivch1}.

The step-like variation of $\hat H$ across the heterointerfaces
results in a singular contribution to the matrix elements
in Eq. (\ref{Schred}) which can be written
as a separate interface term in $\hat H_s$ dependent on
(equivalent to) the choice of boundary conditions:
\begin{equation}
\hat H_s=\mp i\delta (z-z_n)\left(
\begin{array}{cc}
F_1-F_2 & \mathbf{0} \\
\mathbf{0} & F_1^{*}-F_2^{*}
\end{array}
\right).   \label{H-s}
\end{equation}
Here $z_n$ corresponds to the plane of the $n$-th interface,
the sign in the right side of Eq. (\ref{H-s}) is different for
the left and right interfaces of the QW,
and $F_{i}$ in each layer is given by
\begin{equation}
F=\left(
\begin{array}{cccc}
A\hat k_z & 0 & 0 & \frac i2P \\
0 & M\hat k_z & 0 & \frac 12k_xN \\
0 & 0 & M\hat k_z & \frac 12k_yN \\
-\frac i2P & \frac 12k_xN & \frac 12k_yN & L\hat k_z
\end{array}
\right).  \label{F}
\end{equation}
Formally these type of terms arise from the action of
$\hat k_z$ on the step-like dependent Kane parameters (see
Appendix, Eq. (\ref{G-r})).

Here we use the symmetrization prescription of Ref. \cite{Bar}
for each matrix element in the SL Hamiltonian (see Appendix).
However, as was pointed out and discussed in Refs.~\cite{Smith,GelMo},
the induced symmetric form for each matrix element is excessive.
The requirements that the SL Hamiltonian be Hermitian and
have the symmetry of the bulk
materials does not fix the order of operators $%
k_z=-i\partial /\partial z$ and band parameters
in the Hamiltonian. Variation of this order results
in an alteration of the matrix $F$
\begin{equation}
\delta F=\left(
\begin{array}{cccc}
0 & 0 & 0 & \frac{i\alpha }2P \\
0 & 0 & 0 & \frac \beta 2k_xN \\
0 & 0 & 0 & \frac \beta 2k_yN \\
\frac{i\alpha }2P & -\frac \beta 2k_xN & -\frac \beta 2k_yN & 0
\end{array}
\right) ,  \label{delta-F}
\end{equation}
where the parameters $\alpha$ and $\beta$  may vary in the interval
[-1,1] depending on the operator order which can not be rigorously
justified.

The manifestations of $\alpha$ and $\beta $ terms in the SL
spectra are quite different. Since the $\beta $ terms are
proportional to the in-plane components of the wave vector
${\bf k}$ they do not affect the positions of the miniband centers,
but slightly modify the dispersion of valence minibands. These
corrections are of the order of the dispersion of the hh subbands
reduced by the factor proportional to the probability to find
a hole at the interface. In contrast, the $\alpha$ terms shift
the position of the electron- and the lh- miniband centers and
therefore may alter the splitting between the light and heavy hole
minibands.

Note here that the effects of the boundary conditions for the
alloy heterostructures are dependent of the choice of the Kane
Hamiltonian parameters, namely, the evaluation of the first-order
momentum matrix element between the $\Gamma_1$ and $\Gamma_{15}$
states, $|P|$, and the second-order momentum matrix elements and
their difference in the well and barrier layers
\cite{Smith,GelMo}. For the structures with close values of these
parameters in the layers \cite{Ram-Moh} the effects of the
interface in the SL miniband spectra are usually not large.

Then, it should be noted that the effective SL Hamiltonian (with a singular
interface term in the form of Eq. (\ref{H-s})) has the symmetry which follows
from the symmetry of the bulk materials. As a result, for the (100)-oriented
SL structure the reduction of translational symmetry does not lead
to light-heavy hole transformations at the interfaces for holes moving
along the SL axis \cite{Bar}. However, the actual point symmetry
$C_{2v}$ of the (001) interface of the III-V based SL structures allows
the interface-induced mixing between the hh-  and lh- states \cite{Ivch}.
This coupling accounts for the interface-induced optical anisotropy
in the interband transitions and the anisotropic exchange splitting
of the $1s$ heavy-hole exciton level in type II SLs \cite{Ivch1}
and also for the additional spin splitting of the states at $k \ne$ 0
\cite{Voisin}.

The hh-lh mixing at the interfaces results from the mixing between
$X$ and $Y$ orbital states of the valence band edge. This effect
can be taken into account by adding to the valence band part
of the Hamiltonian an extra surface term $H_{v,X-Y}$ \cite{Ivch}
\begin{equation}
H_{v,X-Y}=\sum_i (-1)^i \frac{\hbar ^2t_{x-y}}{m_0a_0}\left[
\begin{array}{cc}
\{I_XI_Y\} & 0 \\
0 & \{I_XI_Y\}
\end{array}
\right] \delta (z-z_i),  \label{Iv1}
\end{equation}
where $a_0$ is the lattice constant, $t_{x-y}$ is a dimensionless
mixing parameter, $z_i$ are the coordinates of the interfaces
and $\{I_XI_Y\}$ is the symmetrized product of the matrixes
for the angular momentum $I=1$,
\begin{equation}
\{I_XI_Y\}=\frac 12\left[
\begin{array}{ccc}
0 & -1 & 0 \\
-1 & 0 & 0 \\
0 & 0 & 0
\end{array}
\right]   \label{Ivch2}
\end{equation}
which is invariant under $C_{2v}$ group operations.
The factors $\pm 1 $ at the right and left interfaces of a
chosen layer provide compensation of the asymmetries
of the two surfaces of the layer in the average.
As a result, at ${\bf k},q $=0 only the hole miniband
states with different parity are mixed.
Then, the heterointerface hh-lh mixing term couples
the hh and lh pairs of states with the angular
momentum projections +3/2 and -1/2 and also -1/2 and +3/2.
The mixing term of Eq. (\ref {Iv1}) should be transformed into
a $8\times 8$ matrix by the inclusion in the matrixes $\{I_XI_Y\}$
the first row and the first column with zero elements
and then added to the SL Hamiltonian
of Eq. (\ref{H-r}).

An important consequence of the lowered symmetry of the
interfaces is the spin splitting of the miniband states
\cite{Voisin,Golub}. Indeed, the SL Hamiltonian (\ref{H-r})
has the inversion center and consequently the band energy
spectrum obtained in Fig. 1 is spin degenerated. The interface
terms (\ref{Iv1}) break this symmetry and
induce the spin splitting linear in $k$ at $k=0$.

Besides, there is also a spin splitting caused by a bulk
inversion asymmetry of each III-V semiconductor layers,
originated from the so-called Dresselhaus terms in the bulk
Hamiltonian. For the electron states both contributions are
negligible due to a small admixture of the hole components to the
conduction band envelopes. However, the spin splitting of the
hole minibands originated from the interface terms (\ref{Iv1})
can be sizable and can exceed the
contribution of the Dresselhaus terms \cite{Voisin}.

The role of the interface originated asymmetry depends on the
value of the mixing parameter $t$ which was estimated for some
heterostructure interfaces in the tight binding approximation
\cite{Ivch,Ivch1,IvchVo}. Here (if not specified differently)
we use for $t$ the interpolated values from Ref. \cite{IvchVo}.
The hole miniband spectra of GaAs/AlGaAs SL allowing $X-Y$ mixing
with $t=0.6$ is shown for ${\bf k} ||$ (110) in the inset of Fig. 5.
The splitting is most noticeable for the hh2
states and is highly anisotropic in the layer plane.

The effect of different boundary conditions and the consequences
of the interface asymmetry for the polarization spectra are discussed
below.

\subsection*{Interband absorption and electron polarization}
The optical absorption coefficient is calculated as
\begin{equation}
\alpha (\omega )=\frac{\left( 2\pi \right) ^{2}e^{2}\omega }{\hbar c}\int
\sum_{n,n^{\prime }} \left|M _{n^{\prime },n}\right| ^{2}
\delta (\omega _{nn^{\prime }}({\bf k}, q)-\omega ) \frac{d{\bf k } dq}
{\left( 2\pi \right) ^{3}},  \label{alfa}
\end{equation}
where $M_{n^{\prime},n}=\left\langle
\Psi_{n^{\prime},\bf{k},q}\left| {\bf er}%
\right| \Psi _{n,{\bf k},q}\right\rangle $ is the dipole matrix element
of the optical transition between the $n$ and $n^{\prime }$ minibands,
$\mathbf{e}$ is the photon polarization vector located in the
interface plane, and $\hbar \omega_{nn^{\prime }}( {\bf k},q)=%
\varepsilon _{n^{\prime }}({\bf k},q)-\varepsilon _{n}({\bf k},q)$ is the
excitation energy. The sum in Eq. (\ref{alfa}) is performed over all
$n$-th occupied  and $n^{\prime }$-th vacant  subbands \cite{excitons}.

In order to calculate the spin polarization of generated electrons
using Eq.(\ref{P}) for the circularly polarized light
propagating along the SL axis
(i.e., for ${\bf e} = ( {\bf e}_x \pm i \ {\bf e}_y)/2$)
one should distinguish the final electron states with "Up" ($\uparrow$)
and "Down" ($\downarrow$) spin projections on this axis.
We will introduce two absorption coefficients
$\alpha_{\uparrow}$ and $\alpha_{\downarrow}$ for the optical
transitions to the $|\uparrow \rangle$ and $|\downarrow \rangle$ final
electronic states, respectively, and assume that
$n_{\uparrow,\downarrow} \propto \alpha_{\uparrow,\downarrow}$.

Then the absorption coefficients, $\alpha_{\uparrow}$ and
$\alpha_{\downarrow}$, are calculated via the density matrix
$\hat{\rho}$ for the optical transitions and the projection
operators $\hat{P}_{\uparrow}$ and $\hat{P}_{\downarrow}$ as
\begin{equation}
\alpha_{\uparrow,\downarrow}=Sp\left\{ \hat{\rho}\hat{P}_{\uparrow,\downarrow}%
\right\} .
\end{equation}
The trace here implies also the integration over ${\bf k}$ and $q$ in
addition to the summation over the electron minibands.
The density matrix for the optical transitions is defined as
\begin{equation}
\begin{array}{c}
\hat{\rho}_{n_{2}n_{1}}\left( {\bf k},q\right) = \\
\frac{\left(2\pi e\right)^{2}\omega }{\hbar c}\sum_{n}M_{n,n_{2}}^{\ast }\left( {\bf k}%
,q\right) M_{n,n_{1}}\left( {\bf k},q\right) \delta (\omega _{nn_{1}}({\bf k}%
,q)-\omega ).  \label{M}
\end{array}
\end{equation}
Here indexes $n_{2}$ and $n_{1}$ correspond to the twofold degenerate
subbband states. The matrix elements of the projection operators are
equal to
\begin{equation}
\left\langle \Psi _{n^{\prime },{\bf k},q}\left| \hat{P}_{\uparrow
,\downarrow }\right| \Psi _{n,{\bf k},q}\right\rangle =C_{\nu}
\sum_{i} A_{i\nu }^{\ast }(n^{\prime },{\bf k},q)A_{i\nu }(n,{\bf k},q),
\end{equation}
where $\nu =1$ for $\hat{P}_{\uparrow }$ and $\nu =5$ for $\hat{P}%
_{\downarrow }$;   $C_{\nu }=\left( \sum_{n,i}\left| A_{i\nu }(n,{\bf k}%
,q)\right| ^{2}\right) ^{-1}$.
The optical matrix elements for the interband transitions in
Eqs. (\ref{alfa},\ref{M}) are calculated straightforwardly as
\begin{equation}
M_{n^{\prime },n}({\bf k}%
,q)=\left\langle \Psi _{n^{\prime },{\bf k},q}\left| {\bf e\nabla }_{{\bf k}}%
\hat{H}\right| \Psi _{n,{\bf k},q}\right\rangle /\hbar \omega _{nn^{\prime
}}({\bf k},q),
\end{equation}
though a singular interface contribution is also present
(e.g. see Ref. \cite{Szmu}).
Squared matrix elements for the transitions to the
$|\uparrow\rangle$  and $|\downarrow\rangle$ subband states
are shown in Fig. 2 as a function of wave vector for the same
structure as in Fig. 1.

\begin{figure}
  \centering
\epsfig{file=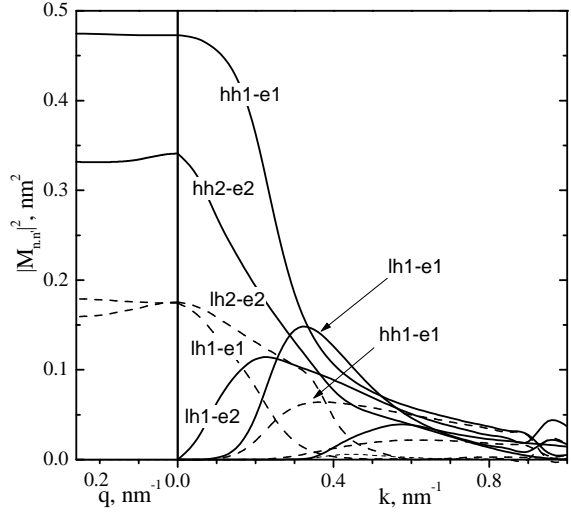,width=7.5cm}
\caption{Squared optical matrix elements for the transitions
to spin-up (solid) and spin down states (dashed lines) as a function
of the wave vector in the plane of the interface ($\bf k ||$ (100))
and normal to the SL interfaces ($q$) for the same structure
as in Fig. 1.}\label{FIG2}
\end{figure}

Strong dispersion of the matrix elements for the interband
transitions under excitation by circularly polarized light
originates from the strict selection rules for the optical
transitions at the boundaries of the Brillouin zone
${\bf k} = 0$ and $q = 0,\pi /d$, connected with the miniband
spin and parity.

Since the Bloch functions of the conduction band ($|S\uparrow,
\downarrow \rangle $) and the valence band
(e.g. $|X\uparrow, \downarrow \rangle $)
have the opposite parity, the optical matrix elements, $M_{n',n}$,
are non-zero for the transitions between minibands with
the same envelope parity (e.g. hh1 $\rightarrow $ e1).
Among the allowed transitions the largest absolute value of
$M_{n',n}$ has transitions between minibands with equal numbers,
e.g. hh1 $\rightarrow $ e1 and lh1 $\rightarrow $ e1,
with the largest overlap of the envelopes.

Besides the spatial dependence, the spin structure of the
envelopes is also important. The spin-polarized electrons
are generated by circularly polarized light with
photon angular momentum $I=1$. For an angular momentum
projection of $+1$, the electron states with
$|\uparrow \rangle$ spin projection on the SL axes
(opposite to the light propagation direction
in the reflection geometry)
will only be populated at ${\bf k} = 0$ due to the transitions
from the hh states with the total angular momentum projection $+3/2$.
All the other optical transitions are forbidden by the
conservation of angular momentum.
Similarly, the $|\downarrow \rangle$ states
are populated by the transitions from the light hole
states with the total angular momentum projection of $+1/2$.
In Fig. 2 we show the squared matrix elements
$\left| M_{n',n}\right| ^{2}$ for main optical transitions,
solid and dashed lines correspond to the transitions
in the  $|\uparrow \rangle$ and   $|\downarrow \rangle$ states,
respectively.

These selection rules are released at ${\bf k}\neq 0$ due to the
mixing of the heavy and light hole components of different parity
and spin projections. The resulting dispersion of matrix elements
in the interface plane for ${\bf k} ||$ (100)
is shown in Fig. 2. With the growth of the in-plane momentum,
$k$, the optical matrix elements decrease for
the allowed transitions and increase for the forbidden
transitions. For example, one can see a rather rapid growth of the
matrix element of the lh1$\rightarrow $e2 transition forbidden
by parity  at ${\bf k} = 0$. It originates from the mixture of
the lh1 miniband with the close hh2 states (see Fig. 1).
Consequently, the oscillator strength of the
lh1$\rightarrow $e2 transition decreases.

\section*{Calculation results}
\subsection*{Unstrained structures}
Up and down contributions to the photoabsorption coefficient
(dashed line and dot-dashed line) and also the total absorption
spectrum (solid line) of the unstrained GaAs/AlGaAs SL are shown
in Fig. 3 (a). To give deeper insight into the optical spectrum
formation the partial contributions of the most important optical
transitions are also shown by thin solid lines.

In a manner similar to the density of states in multi-QW structures,
the optical absorption spectrum has a step function character,
the most noticeable steps
coming from transitions that have large enough optical matrix
elements at their absorption edge. The steep edges in the polarized
light absorption hh1-e1 and hh2-e2 are responsible for the steps
in the generation of $|\uparrow \rangle$  polarized electrons,
while the transitions lh1-e1 and lh2-e2 provide the steps
in the $|\downarrow \rangle$ electron excitation.

The transition e2-hh2 needs more comment. According to Fig. 3 (a)
the contribution of the hh2 - e2 transition itself provides only
half of the second step in up-polarized absorbtion. The other
half is the contribution of the e2 - lh1 transition which is
forbidden by parity at the absorption edge and therefore should
not contribute at all. However, in the SL structure under
consideration the edges of hh2 and lh1 minibands are very close.
This fact leads to a strong mixing between these states resulting
in the sharing of the oscillator strength between the hh2-e2 and
lh1-e1 transitions.
\begin{figure}
\centering \epsfig{file=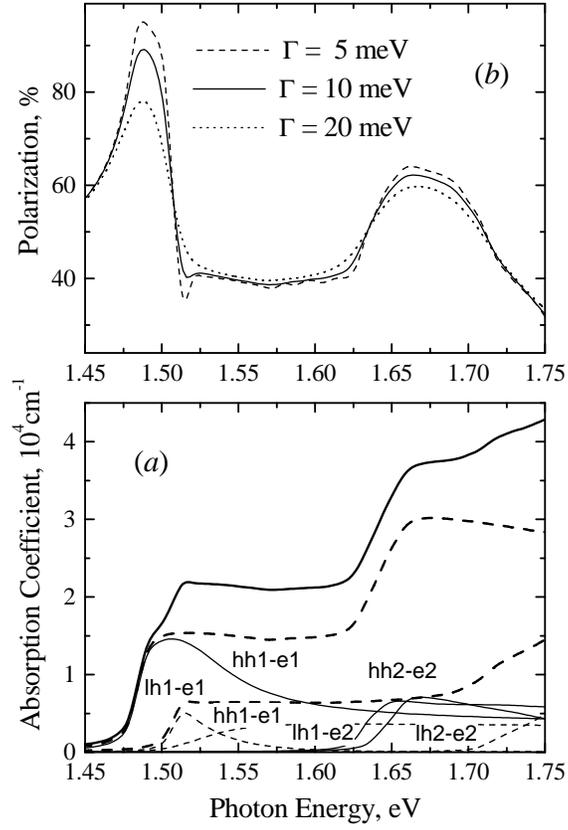,width=7.4cm} \caption{Calculated optical absorbtion
of GaAs/Al$_{0.4}$Ga$_{0.6}$ As superlattice with $a$=8 nm, $b$=4nm and $\gamma =7$
meV (a) and polarization spectrum of this superlattice (b); polarization spectra
calculated for smaller ($\gamma $= 5 meV) and larger ($\gamma $= 20 meV) level width
broadening are shown by dashed and dotted lines.}\label{FIG3}
\end{figure}

While the overall shape of the polarized absorption spectrum is mostly
determined by the optical transitions between the electron and hole
subbands with equal quantum numbers, similarly to the vertical optical
transitions in the bulk, we note that these transitions are responsible
only for the sharp step-like edges in the optical spectrum, their
positions and magnitudes.  The contribution of these transitions
decreases rapidly away from their edges. This feature results from
the decrease of the optical matrix elements with the growth of
the electron momentum in the heterostructure plane \cite{ADA}
(see Fig. 2). The rather flat steps in the absorption come from
the combined contribution of different optical transitions.

For example, the allowed lh1-e1 transition gives rise to the
first step edge for the down-polarized absorption, though an
almost constant value of generation rate at larger photon
energies results from the additional contribution of the hh1-e1
transition. Note here that the hh1-e1 transition does not
contribute at its edge to the generation of down polarized
electrons at all.

The step structure of the absorption spectrum determines the
excitation energy dependence of the electron polarization at the
excitation shown in Fig. 3 (b). In the interval between the left
step edges in up and down absorption rates the electron
polarization is close to 100 \%. Similarly to the strain-induced
electron polarization, this effect arises from the splitting
between the heavy and light hole subbands in the SL structure.
The width of the first polarization peak depends on the energy
separation between hh1 and lh1 minibands. However, in addition to
the first peak, there are also the repetitions of the first
polarization peak associated with the higher electronic minibands.
The strongest is the second polarization peak (see Fig. 3 (b)),
which originates from the optical transitions to the second
electron miniband, e2. Its position, shape and the maximum
polarization value are dependent on the dispersion of the
e2- miniband and the resulting optical density of states
while its width is determined by the distance between the hh2 and
lh2 levels. Therefore, the polarization magnitude in the second
polarization peak can exceed $P$ = 50 \% (the maximum value for
bulk crystal). Then polarization decreases with the growth of the
excitation energy and total absorption value. This decrease
becomes very rapid when excitation energy is enough to induce
transitions from the spin-orbital split valence band.

Note, that the shape of the polarization curve and the maximum
polarization are sensitive to the width of the miniband levels and
to the smearing of the interband absorption edges.
In our calculations we use the Lorentzian broadening of the quantum
levels. The actual level width in the real heterostructure originates
from the interaction of the carriers with phonons and impurities and
from fluctuations of layer composition and thickness
and possibly other heterostructure defects. For the best SL
structures absorbtion edge smearing was reported to be
in the range  $\gamma $ =7-12 meV at room temperature \cite{Tails},
depending on doping and structural quality.
Within the present work we use the level width as a parameter.

In Fig. 3 (a) the absorption  and electron polarization spectra
calculated for  $\gamma $ =10 meV are shown by the solid lines.
The variation of the polarization spectrum with the level width
is shown in Fig. 3 (b) by the dashed (for $\gamma $ = 5 meV)
and dotted ($\gamma $= 20 meV) lines. For larger values of
$\gamma $ along with the smearing of the absorption step edges,
the polarization peaks become more narrow and smaller in magnitude.
The long tail in the lh - e1 optical transitions reduces
the amplitude of the main polarization maximum.
Thus, the broadening of the SL energy spectrum
sets the limit for the highest possible initial
photoelectron polarization. A destructive effect of the
broadening is reduced in the case of a wider
polarization peak, i.e. in the case of a larger
hh - lh miniband splitting.

In Fig. 4 the SL polarization spectra for different barrier
thickness $b$ = 1.5, 2.5 and 4 nm are depicted to show the
influence of tunneling through the SL barriers. For the barriers
thicker than 4 nm the polarization spectra are found to be
similar to that of the MQW structure with nontransparent
barriers. For thinner barriers the dispersion of the minibands
along the SL axis starts to result in a narrowing of the
polarization peaks along with the decrease of the polarization.
\begin{figure}[h]
\centering
\epsfig{file=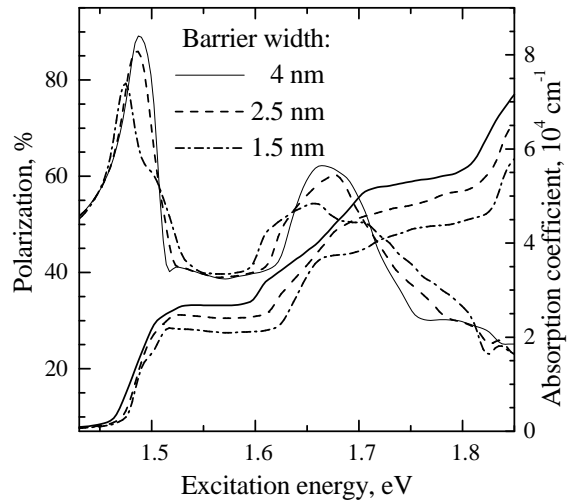,width=7.4cm}
\caption{Polarization and absorption spectra for
the GaAs/Al$_{0.4}$Ga$_{0.6}$As SL structures with $a$ = 8 nm
and different barrier layer thicknesses; $\gamma = 10$ meV.}
\label{FIG4}
\end{figure}
This decrease is more pronounced on the right side of the first
polarization maximum showing the effect of the transitions from
the lh minibands. The tunnelling probability of light holes is
larger than that for heavy holes which results in a larger
dispersion along the SL axis and therefore a larger spread of
their input to the polarization spectrum. This dispersion is also
manifested in the smearing of the high-energy features in the
absorption spectra of Fig. 4. In the limit of extremely narrow
barriers $b \le$ 1 nm, the peak structure is damped while the SL
polarization spectrum is transformed into the spectrum of the bulk
material.

It should be noted that the width of the first polarization peak
due to the splitting of the lh1 and hh1 minibands in unstrained
SL structures with thick QW layers can be estimated as $\Delta%
E_{\rm hh1, lh1} \approx \hbar^2 \pi^2 /a^2(m_{\rm lh}^{-1}- %
m_{\rm hh}^{-1})$ (i.e. the value for an isolated deep QW) and is
enlarged in the structures with narrowed wells. However
the rapid growth of the splitting $\Delta%
E_{\rm hh1, lh1} \propto a^{-2}$ with a decrease of $a$ presents itself only when
$E_{lh1} \ll \Lambda_v$ where $E_{lh1}$ is the lh1 miniband energy at ${\bf k},q=0$.
Then the splitting saturates and decreases, mainly due to the effects of lh tunnelling
in the barriers.

Actually, the maximum valence band splitting is obtained for the structures where
$E_{\bf lh1}$ is substantially smaller than $\Lambda_{\rm v}$. For the
GaAs/Al$_{0.4}$Ga$_{0.6}$As SL it turns out to be about four times smaller than
$\Lambda_{\rm v}$. The maximum splitting value 43 meV is achieved at the well width
$a$ = 2.6 nm. The calculated polarization spectrum of this SL with $a$ = 2.6 nm, $b$ =
3.6 nm is presented in Fig. 5 showing the growth of the maximum polarization value to
$ P \approx $ 93 \% at the maximum hh1-lh1 splitting.
\begin{figure}[h]
  \centering
 \epsfig{file=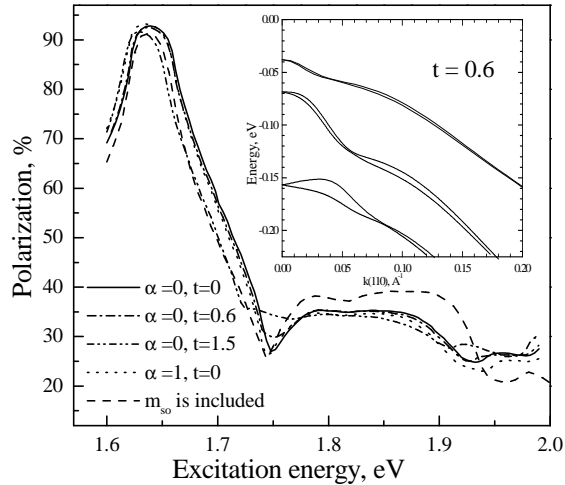,width=7.4cm}
 \caption{The influence of the interface hh-lh band mixing ($t$
 being the mixing parameter) on the polarization spectrum of
 InGaAs/AlGaAs structure for two choices of the boundary
 conditions ($\alpha$). Dashed line shows results of the
 calculations with interband matrix element $P$ determined with
 the use of effective mass of the spin-orbit split valence band;
 $\gamma = 10$ meV; the hole
miniband spectra is shown in the inset.}\label{FIG5}
\end{figure}
Finally, the splitting is reduced in structures with thin
valence band barriers, transparent for light holes, while in
structures with high and low-transparent barriers in the
conduction band the electrons are not mobile enough. This
makes unstrained structures with typically higher barriers
for electrons less favorable for polarized photoemission.

\subsection*{Interface effects}
As stated above, the choice of the boundary conditions
determined by the behavior of the SL Hamiltonian
in the heterointerface regions, can result in variations in
the calculated absorption and polarization spectra and in
the maximum polarization value. These variations are
illustrated by Fig. 5 in which the polarization
spectra for 2.8 x 3.6 nm GaAs/Al$_{0.4}$Ga$_{0.6}$As SL
with different boundary conditions are depicted.

The spectra obtained for various assumptions about $\alpha$
values are rather close, the deviations between
the $\alpha =0$ and $\alpha =1$ cases being essentially due to
a small shift of the electron quantum levels. The influence of
the $\beta $ terms on polarization spectra is found
to be negligible. Since the boundary conditions can not be
derived unambiguously within the effective mass approximation
the deviations in the spectra can be regarded
as an error bar for the approach employed and in what follows
we take $\alpha = \beta=0$.

The minor effects of the boundary conditions on the calculated
spectra result from the fact that the first-order momentum
matrix elements $P$ do not differ much in the SL layers.
The parameters of the effective mass Hamiltonian (\ref{H-r}),
including $P$, can be defined in terms of the experimental
band gap, spin-orbital splitting,
effective masses, and the $E_{\bf p}$ values, $E_{\bf p} =2 m_0
P^2/\hbar^2$, evaluated from the experimental data on the
nonparabolicity of the conduction band spectrum \cite{Ram-Moh}.
In our calculations we use for the Hamiltonian parameters
interpolation scheme and values recommended in Ref.
\cite{Ram-Moh}.

The alternative is to use the data on the spin-orbit split-off
band effective mass but to ignore the $E_{\bf p}$ data, see a
dashed-line spectrum in Fig. 5. In this case the calculation
results are considerably more sensitive to the choice of the
boundary conditions which makes the conclusions that could be
drawn from them less reliable.

Then, different interface effects originate from the additional
term (\ref{Iv1}) in the SL Hamiltonian which mixes hh
and lh states. In Fig. 5 we show by the dotted line the polarization
spectrum of the GaAs/Al$_{0.4}$Ga$_{0.4}$As SL allowing for
the interface hh-lh mixing. The value of mixing
parameter $t=0.6$ is taken by an interpolation of
the values given in Ref. \cite{IvchVo}.
Note here that since the term couple lh and hh states
with the opposite parity, the major effect comes from the
hh2 and lh1 miniband interaction. While the mixing is neglected,
the repulsion between closely situated hh- and lh- minibands
at $k \neq 0$ leads to a large or even negative effective mass
of the upper subband manifested as an additional feature
in the absorbtion and a dip in polarization spectra.
The repulsion between lh1 and hh2 minibands at the
zone center increases the interlevel distance and therefore
smears the described feature in the polarization spectrum.

The linear in $k$ anisotropic splitting of the hole minibands
(seen in the inset of Fig. 5) being integrated over the $k$
directions does not show itself in polarization spectra.

The influence of the interface term of Eq. (\ref{Iv1})
on the main polarization maximum is of minor importance
for $t \le$ 1. The admixture of the lh2 component to the
hh1 wave function does not affect the amplitude
of the polarization peak since it does not contribute
to the optical transition to the e1 electron
state with an opposite parity, the resulting shift of
the hh1 miniband being not larger than a few meV.

\subsection*{Strained structures}
In strained SL structures the lattice mismatch between SL layers
and a substrate is aimed to results in an additional hh- lh-
miniband splitting. Several types of structures favorable
for photoemission with (i) strained SL wells \cite{Nakani,Maru}
(ii) stained barriers \cite{ArSu,Ambra} and
(iii) employing strain compensation \cite{nano03}
are designed and tested. We now discuss how polarization spectra
evolve with strain for realistic SL structures.

(i) {\it Strained wells}. First, we consider polarization spectra
of a (100)-oriented GaP$_{0.4}$As$_{0.6}$ 4x4 nm SL structure
with strained well layers. The band offsets in this SL are
comparable to these in the Al$_{0.4}$Ga$_{0.6}$As SL structure
but the lattice constant in the barriers is smaller than that in
the wells. The resulting deformation of the SL with finite total
thickness strongly depends upon the substrate lattice constant.
In Fig. 6 we show the optical absorption and polarization spectra
for GaAs/GaP$_{0.4}$As$_{0.6}$ SL calculated for the structure
grown on GaP$_{\rm{x}}$As$_{\rm{1-x}}$ pseudo-substrates with
three different concentrations of P. The thickness of the SL
structure is assumed to be much less than that of the substrate
so that the resulting lattice constant of the layers in the layer
plane accommodates completely to the substrate.
\begin{figure}[h]
\centering
\epsfig{file=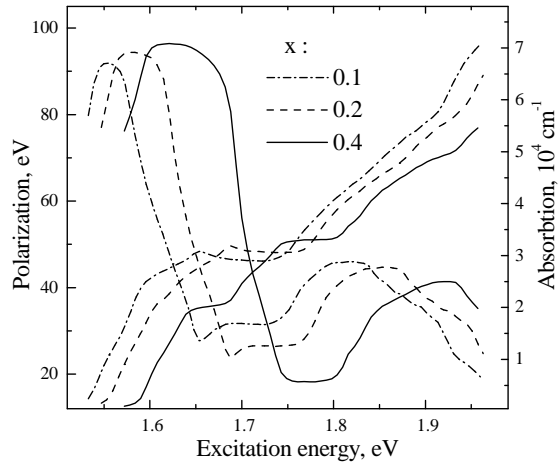,width=7.4cm}
\caption{Polarization and absorption spectra of strained
well GaAs/GaAs$_{0.6}$P$_{0.4}$ SL structure
for different P concentrations, $x$, in the substrate;
$\gamma = 10$ meV.}\label{FIG6}
\end{figure}
The GaAs QW layers are stressed in the layer plane so that the
hh- subband is pushed up while the lh- subband moves down.
The strain in the QW layers and therefore the valence splitting
in QW increases with the growth of the P concentration in the
substrate. For the substrate compositions considered,
x=0.1, 0.2, 0.4, the hh-lh valence band splitting is equal
to 23, 44 and 82 meV, respectively. Since the barriers are
tensile strained in the case of x=0.1, 0.2 and unstrained
at x=0.4, the shifts of the barrier lh- and hh- band edges
due to the barrier deformation has an opposite sign. As a
result the valence band offset for the light holes
($\Lambda_{lh}=0.18$ eV) becomes smaller than for the heavy
holes ($\Lambda_{hh}=0.28$ eV), their values being
almost independent upon the substrate composition.

The decrease of the well depth for the light holes reduces the
difference between the confinement energies of hh1 and lh1
miniband states at ${\bf k},q$ =0 to about 20 meV which is almost
two times smaller than that in the unstrained
GaAs/Al$_{0.4}$Ga$_{0.6}$As SL with the same layer width. The
total energy splitting between hh1 and lh1 states, being the sum
of the confinement energy difference and the QW valence band
splitting, equals to 45, 65 and 99 meV (for x = 0.1, 0.2 and 0.4,
respectively).

Fig. 6 shows the evolution of the spectra with the increase of
strain in the well layers. It can be seen that the steps in
absorption spectra and the main polarization peak in the strained
SL become wider as a result of the larger hh1 and lh1 miniband
separation due to join effects of the confinement and strain. The
maximum polarization value rises to $\approx$ 97 \% for a SL with
stressed wells and unstrained barriers (i.e. for $x=0.4$), while
the minimum in the polarization spectra becomes more deep and
wider.

(ii) {\it Strained barriers}. Choice of the barrier and well
compositions allows one to design SL structures with strained
barriers in which the valence band splitting increases the
barrier height for light holes and decreases it for the heavy
holes. As a result the difference between the lh- and hh- state
energies is enlarged.

As an example we consider the polarization spectra for a GaAs
based structure with ternary Al$_{\rm x}$ In$_{\rm y}$ Ga$_{\rm
1- x -y}$ As alloy barriers grown on a GaAs substrate. The
lattice constant of the barrier alloy is larger than in GaAs,
which provides the needed sign of the valence band splitting in
the barriers, while the QW layers remain unstrained. The barrier
alloy composition Al$_{0.21}$ In$_{0.2}$ Ga$_{0.6}$ As is chosen
to achieve a close to zero value of the conduction band offset
and therefore a high vertical electron mobility appropriate for
photoemission \cite{MamCle,ArSu}.
\begin{figure}[h]
\centering \epsfig{file=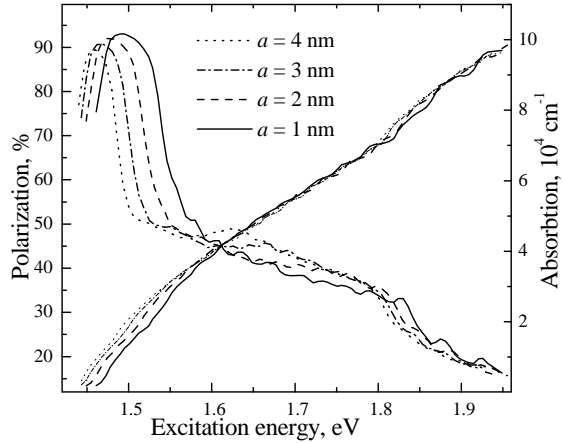,width=7.4cm} \caption{Polarization and absorption
spectra of GaAs/Al$_{0.21}$ In$_{0.2}$Ga$_{0.69}$ As SL with strained barriers and
close to zero conduction band offset with $b$ = 4 nm and different QW thicknesses,
$\gamma = 10$ meV.}\label{FIG7}
\end{figure}
In Fig. 7 the optical absorption and polarization spectra
calculated for this type of SL with 4 nm barriers and different
well thicknesses are shown. The valence band offsets for hh- and
lh- subbands are equal to 62 and 140 meV, respectively.
Consequently the energy difference between the hh1 and lh1
miniband states becomes larger than that in the GaAs QW SL with
unstrained 100 meV- hight barriers though the hh1-lh1 splitting
remains smaller than the deformation splitting in the barrier
layers. Small conduction band offset results in the complete
smearing of the features in the absorbtion spectra. A high
polarization ($\approx$ 92 \%) is obtained in the structures with
considerably thinner ($\le$ 2 nm) QW layers with hh1-lh1
splitting of about 64 meV.

(iii) {\it Strain compensation.} The thickness of a strained SL necessary for
photoemission exceeds by an order of magnitude the critical thickness for the coherent
layer growth, which leads to the strain relaxation, defective structures and
polarization losses. The use of strain compensation, whereby the composition of the SL
barrier layers is chosen to have opposite (tensile) strain from that of the quantum
well layers is proposed to overcome limitations on the overall thickness of the SL
structure \cite{nano03}.

The calculated spectra of strain compensated
In$_{0.2}$Ga$_{0.8}$As/GaAs$_{0.65}$P$_{0.35}$ Sl structures on
GaAs substrate with different well width are shown in Fig. 8. In
this structure the valence band splitting in the well and barrier
layers have a close magnitude $\approx 78 meV $ but differ by the
sign. Consequently the well depth for the heavy holes is five
times larger than for the light holes. Thus the heavy hole
confinement energy has much stronger dependence on the well
thickness then the lh1 energy and therefore the hh-lh splitting
is mainly determined by the position of the hh1 miniband. The
maximum polarization is obtained for the structures with thicker
QW ($a \ge$ 4 nm) layers, with hh1-lh1 splitting ($\ge$ 60 meV)
remaining lower than splitting caused by deformation in QW
layers. It should be noted that the splitting is considerably
reduced in the structures with thin QW layers due to more rapid
growth of the hh1 state energy.
\begin{figure}[h]
\centering
\epsfig{file=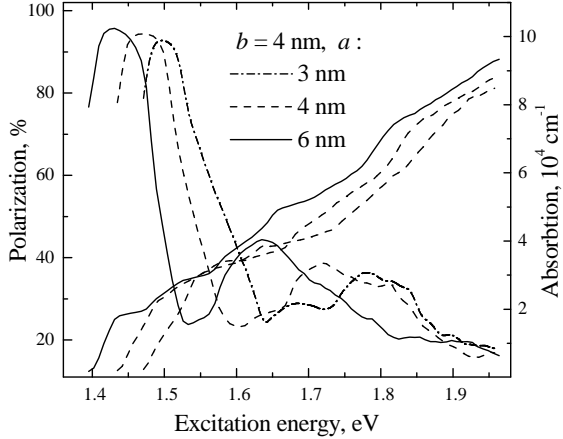,width=7.4cm}
\caption{Polarization and
absorption spectra of strain-compensated Al$_{0.1}$
In$_{0.2}$Ga$_{0.7}$ As/GaAs$_{0.7}$ P$_{0.3}$ SL with $b$ = 4 nm
and different QW thicknesses; $\gamma = 10$ meV.}\label{FIG8}
\end{figure}
Note here that the resulting valence band
splitting and the electron initial polarization in the strain
compensated structures are lowered (compared to strained well
structures) due to the lowered barriers for the light holes
formed by the tensile-strained layers.

A drawback of the strained well and strain compensated structures
is typically the high barriers for electrons having reduced
transparency, a disadvantage which is absent in strained barrier
SL with zero conduction band offsets.

\section*{Discussion}
Calculated dependencies of the electronic polarization of the excitation
energy are close to the experimental spectra of polarized electron emission
for investigated structures \cite{MamCle,Nakani,Maru,Ambra}. The value of
polarization in the first polarization maximum in the SL structures under
consideration ranges from 87 to 95\%
for the broadening of the absorption edge $\gamma$=10 meV taken here, and is
noticeably growing with the splitting between the hh- and lh- states of the valence
band. We have shown that the strain and confinement effects on the hh1- lh1 splittings
do not simply add up mainly because the lh confinement energy is strongly influenced
by the tunnelling in the barriers, the range of optimal layer thickness being
different for the strained QW and strained barrier structures.

The actual value of the relevant broadening parameter $\gamma$ is
sensitive to all the processes leading to the structurally
induced tails in the interband absorption but is not identical
to the energy width of the absorbtion tail since it
relies on the contribution of the lh-e transitions to
the absorption at the hh-e absorption edge.
Therefore, the broadening should also depend on the lh- hh-
splitting. A large separation of the hh1- and lh1- valence
band states in the strained-well SLs make them the most
advantageous for polarized photoemission.

Then, the broadening effects may not be strong enough to account for large
polarization losses observed in some of the strained SL structures
\cite{Nakani,Maru,nano03} within a limited variation of the SL parameters. Additional
polarization losses can be caused by the mixing between lh- and hh- states at the
interfaces due to the lowered symmetry of the interface. For the case of ideally
perfect interfaces discussed above this mixing does not affect the maximum
polarization value in the first polarization maximum due to the left- and
right-interface compensation. In the case of a real interface and allowing
fluctuations of the alloy composition in the interface plane, the left- and right-
interface contributions to the hh-lh mixing will not compensate each other. These
effects are of a particular importance in the structures with "no common anion"
heterointerfaces with enlarged value of the mixing parameter $t$ \cite{Voisin,IvchVo}.

Finally, the polarization of the electrons emitted in vacuum
is also reduced in transport to the surface and emission through
the surface band bending region and the activation
layer. Time and energy-resolved experimental studies of the
polarized photoemission give the estimates of these effects.
Though the polarization losses in transport are not dominating,
the mechanisms and relative contributions of the depolarization
at these stages are still under debates \cite{Kurt}. The set of
wavefunctions and the SL band spectra obtained here are prerequisites
for calculations of the transport and spin relaxation parameters.

\section*{Conclusions}

To summarize, a theoretical investigation of the electronic optical orientation in the
unstrained, well-strained and the barrier-strained superlattices showed the dominating
effect of the valence band splitting on the maximum polarization that can be achieved
in the moment of excitation. The maximum splitting is found in the strained well
superlattices, while the splitting in strain-barrier and strain-compensated structures
with realistic parameters is lower that the deformation induced splitting. However,
the achievable splitting values in these SLs can still be larger than in structures
with one strained layer due to reduced strain relaxation effects.

The highest value of the polarization in the excitation moment is obtained for the
structures where a high strain valence band splitting in the QW layers is accompanied
by a large confinement splitting due to high barriers in the valence band.
Technological limitations for the growth of the highly strained SL structures set
restrictions to the realistic choice of the structures parameters, necessitating
optimization of the photoemitter structure as a whole within these limitations.

\subsection*{Acknowledgments}
This work was supported by CRDF under grant RP1-2345-ST-02, SNSF
under grant 7IP 062585 and RFBR under grant 00-02-16775. We are
grateful to J.E. Clendenin for reading the manuscript.

\section{Appendix}

Matrixes  $G$ and $\Gamma $ in the effective Hamiltonian
(\ref{H-r})  are defined e.g. in Ref. \cite{Kane}.
\begin{equation}
\Gamma =-\frac \Delta 3\left(
\begin{array}{cccc}
0 & 0 & 0 & 0 \\
0 & 0 & 0 & -1 \\
0 & 0 & 0 & i \\
0 & 1 & -i & 0
\end{array}
\right) ,  \label{Gamma-r}
\end{equation}

\begin{equation}
G = \\
\left(
\begin{array}{cccc}
F_c & iPk_x & iPk_y & \frac i2\left( P\hat
k_z+\hat k_zP\right)  \\
-iPk_x & F_x & Nk_xk_y-i\frac \Delta 3 &
\frac 12k_x\left( N\hat k_z+\hat k_zN\right)  \\
-iPk_y & Nk_xk_y+i\frac \Delta 3 & F_y&
\frac 12k_y\left( N\hat k_z+\hat k_zN\right)  \\
-\frac i2\left( P\hat k_z+\hat k_zP\right)  & \frac 12k_x\left( N\hat
k_z+\hat k_zN\right)  & \frac 12k_y\left( N\hat k_z+\hat k_zN\right)  & F_z
\end{array}
\right)   \label{G-r}
\end{equation}

Here $\hat k_z=-i\partial /\partial z$, $k_x$ and $k_y$ are components of
the momentum $\mathbf{k}$,
\begin{equation}
F_c = E_c+A(k_y^2+k_x^2)+\hat k_zA\hat k_z
\end{equation}
\begin{equation}
F_x =  E_v+Lk_x^2+Mk_y^2+\hat k_zM\hat k_z
\end{equation}
\begin{equation}
F_y =  E_v+Lk_y^2+Mk_x^2+\hat k_zM\hat k_z
\end{equation}
\begin{equation}
F_z = E_v+\hat k_zL\hat k_z+M(k_y^2+k_x^2)
\end{equation}
Inside the SL layers 1 and 2 all parameters in (%
\ref{Gamma-r},\ref{G-r})  $E_{c,v}$,$\Delta $,$P$,$A,L,M,N$ are constant
and equal to the corresponding values of the bulk materials. Therefore
the order of the operators $\hat k_z$ and Hamiltonian parameters is not
essential. However in the interface region where parameters are considered
as step functions this order becomes important. We use the symmetrized
form following \cite{Bar}.


\begin{thebibliography}{99}

\bibitem{Slac}  K. Abe et al. Phys. Rev. Lett. {\bf 86} 1162
(2001).

\bibitem{JLab} M. Poelker, P. Adderley, J. Clark, A. Day, J. Grames,
C. Sinclair et al., {\em Proc 14th Intern Spin Physics Symp.
(SPIN 2000)} (AIP Conf. Proc. Vol. 570, Eds. K. Hatanaka et al.,
Melville, NY) p. 698.

\bibitem{Mami}  K. Aulenbacher, Ch. Nachtigall, H.G. Andersen,
P. Dresher, H. Euteneur, H. Fischer, D. v. Harrach, P. Hartmann,
J. Hoffman, P. Jennewein, K.-H. Kaiser, H. J. Kreidel, S.Plutzer,
E. Reichert, K.-H. Steffens, and M. Steigerwald, Nucl. Instrum.
and Methods Phys. Res. A {\bf 391}, 498 (1997).

\bibitem{SPIN2000} {\em Proc 15th Intern Spin Physics Symp.(SPIN 2002)}
(AIP Conf. Proc. Vol. 570, Eds. K. Hatanaka et al., Melville, NY) p. 698.

\bibitem{OptOr}  {\em Optical orientation,} edited by F.Meier and
B.P.Zakharchenya, North-Holland, 1984.

\bibitem{Mair}  R.A.Mair, R. Prepost, E.L. Garvin, and T.Maruyama, Phys.
Lett. A, {\bf239}, 277 (1998).

\bibitem{SuKal}  A.V. Subashiev, Yu.A. Mamaev, B.D. Oskotskii, Yu.P. Yashin,
and V.K. Kalevich, Semiconductors {\bf33}, 1182 (1999).

\bibitem{Kurt}  K. Aulenbacher, J. Schuler, D. v. Harrach, E. Reichert,
J. Roethgen, A. Subashiev, V. Tioukine, and Y. Yashin, J. Appl. Phys.
{\bf 92}, 7536 (2002).

\bibitem{ADA}  A.D. Andreev, and A. V. Subashiev, Physica E, {\bf13},
556 (2002).

\bibitem{MamCle}  A.V. Subashiev, A.V., Mamaev, Yu.A., Yashin, Yu.P.,
Clendenin, J.E., {\em Phys. Low-Dim. Structures}, {\bf 1/2}, 1 (1999).

\bibitem{Nakani}  T. Nakanishi, H. Aoyagi, T. Kosugoh, S. Nakamura,
M.Tawada, H. Horinaka, and Y. Kamiya, Surface Science, {\bf 454-456},
1042 (2000).

\bibitem{CHerm}  A. Twardowski, C. Hermann, Phys. Rev. B {\bf 43}, 8144
(1987).

\bibitem{Merk}  I.A.~Merkulov, V.I.~Perel', M.E.~Portnoi, Zh. Exp. Teor.
Fiz. {\bf 99}, 1202 (1990) [Sov. Phys. JETP {\bf 72}, 669  (1991)].


\bibitem{Bar}  G.A. Baraff and D. Gershoni, Phys. Rev. B {\bf 43}, 4011
(1991), D. Gershoni, C. H. ~Henry and G.A. Baraf, IEEE Journ.
Quant. Electron., {\bf 29} 2433 (1993).

\bibitem{Kane}  E. O. Kane, ''Energy Band Theory'', in \emph{Handbook on
Semiconductors}, vol. 1, ed. W. Paul, North-Holland, 1982. pp. 193-217.

\bibitem{Ram-Moh}  I. Vurgaftman, J. R. Meyer and
L.R. Ram-Mohan, J. Appl. Phys. {\bf 89}, 5815 (2001).

\bibitem{Smith}  D. L. Smith and C. Mailhiot, Rev. Modern Phys., {\bf 62}
173 (1990).

\bibitem{IvPi}E. L. Ivchenko and G.E. Pikus, {\em Superlattices
and Other Heterostructures} (Springer, Berlin, 1995).

\bibitem{GelMo} A. V. Rodina, A. Yu. Alekseev, Al. L. Efros, M. Rosen and
B.K. Meyer,  Phys. Rev. B {\bf 65}, 125302-1 (2002).

\bibitem{Ivch} E.L. Ivchenko, A.Yu. Kaminski, and I.L. Aleiner,
Zh. Exp. Teor Fiz. {\bf 104}, 3401 (1993) [JETP  {\bf 77}, 609 (1993)].

\bibitem{Ivch1}  E.L. Ivchenko, A.Yu. Kaminski, and U. R\"ossler, Phys.
Rev. B {\bf 54}, 5852 (1996).

\bibitem{Voisin} L. Vervoort, F Ferreira, P. Voisin, Phys. Rev.
B {\bf56},  R12 744 (1997).

\bibitem{Golub}  L.E. Golub, Phys. Rev, B {\bf 56},  R12 744 (1997).

\bibitem{IvchVo} E.L. Ivchenko, A.A. Toropov, and P. Voisin,
Fiz.  Tver. Tela {\bf40}, 1925 (1998)
[Solid State Physics,  {\bf 77}, 609 (1993)].


\bibitem{excitons}  Excitonic effects in the absorption spectra at the
excitation edges are not included since they do not give noticeable
cobntribution to the photoemission current and no signes of these
effects in photoemission were observed experimentally for the presence.

\bibitem{Szmu} F. Szmulowicz, Phys. Rev. B {\bf 51} 1613 (1995).

\bibitem{Tails} C.W. Greef, H. R. Glyde, Phys. Rev. B {\bf 51}, 1778
(1995).

\bibitem{Maru}  T. Maruyama, {\em Proc. of the 14-th Intern. Symp. on
High-Energy Spin Physics, (SPIN 2002)} (AIP, Melville, New York, 2003) 1029.

\bibitem{ArSu} A. V. Subashiev, Yu. A. Mamaev, Yu. P.Yashin,
V. M. Ustinov, A. E. Zhukov, J. E. Clendenin, T. Maruyama, G. A. Mulhollan,
SLAC-Pub 7922, 1998, also in {\em Proc. of the 24-th Intern. Conf. on
Physics of Semicond}, (ICPS-24).

\bibitem{Ambra} A. N. Ambrajei, J. E. Clendenin,
A. Yu. Egorov, Yu. A. Mamaev, T. Maruyama, G. A. Mulhollan,
A. V. Subashiev, Yu. P. Yashin, V. M. Ustinov, A. E. Zhukov,
Appl. Surface Science,  {\bf 166}, 40 (2000).

\bibitem{nano03} Yu. P. Yashin, J. S. Roberts,
L. G. Gerchikov, P. A. Houston, A. N. Ipatov,  Yu. A. Mamaev, A. V. Rochansky
and A. V. Subashiev {\em 11th Int. Symp. "Nanostructures: Physics
and Technology" (NANO 2003)} (St.Petersburg, 2003) p. 61.

\end{thebibliography}
\end{document}